# A cost-benefit *'source-receptor'* framework for implementation of Blue-Green flood risk management

**Christos Iliadis** [1, *], **Vassilis Glenis** [1] **and Chris Kilsby** [1,2]

[1] School of Engineering, Newcastle University, Newcastle Upon Tyne, NE1 7RU, UK;
[2] Willis Research Network, 51 Lime St., London, EC3M 7DQ, UK;
[*] Correspondence: c.iliadis2@newcastle.ac.uk;

## Highlights

- Identifies locations contributing to surface water flooding (sources).

- Identifies buildings and locations at high flood risk (receptors).

- Outlines the cost-benefit nexus between the *'source'* and the *'receptor'*.

- Identifies priority options to mitigate flooding at the *'receptor'* by adding Blue-Green Infrastructure (BGI) in critical locations.

- Combination of extreme rainfall information, flood dynamics and the cost-benefits in an urban area.

**Abstract:** As floods are a major and growing source of risk in urban areas, there is a necessity to improve flood risk management frameworks and civil protection through planning interventions that modify surface flow pathways and introduce storage. Despite the complexity of densely urbanised areas (topography, buildings, green spaces, roads), modern flood models can represent urban features and flow characteristics in order to help researchers, local authorities, and insurance companies to develop and improve efficient flood risk frameworks to achieve resilience in cities. A cost-benefit driven *'source-receptor'* flood risk framework is developed in this study to identify (1) locations contributing to surface flooding (sources), (2) buildings and locations at high flood risk (receptors), (3) the cost-benefit nexus between the *'source'* and the *'receptor'*, and finally (4) ways to mitigate flooding at the *'receptor'* by adding Blue-Green Infrastructure (BGI) in critical locations. The analysis is based on five steps to identify the *'source'* and the *'receptor'* in a study area based on the flood exposure of buildings, damages arising from flooding and available green spaces with the best potential to add sustainable and resilient solutions to reduce flooding. The framework was developed using the detailed hydrodynamic model CityCAT in a case study of the city centre of Newcastle upon Tyne, UK.

The novelty of this analysis is that firstly, multiple storm magnitudes (i.e. small and large floods) are used combined with a method to locate the areas and the buildings at flood risk and a prioritized set of best places to add interventions upstream and downstream. Secondly, planning decisions are informed by considering the benefit from reduced damages to properties and the cost to construct resilient BGI options rather than a restricted hydraulic analysis considering only flood depths and storages in isolation from real world economics.

## 1. Introduction

Surface water flooding is a major and increasing hazard in cities, where assets, properties, and humans are directly affected. The extent and severity of the damage caused by urban floods, around $40 billion per year according to OECD (2016), is a product of both the intensity and the duration of a storm and its interaction with the complex flow paths of a city on the surface and below ground. Extreme storm events are expected to increase worldwide and therefore constitute a critical issue in flood risk analysis and the design or management of critical adaptation solutions is a necessity to reduce and, in some cases, control the flow and the volume of flood water in the urban fabric (Galiatsatou & Iliadis, 2022).

It is expected that by 2050 almost 75% of the world's population will live in urban areas (Liu et al., 2014). The combination of climate change and the increasing urbanisation with the frequency of storms will lead dense areas to improve the current flood mitigation strategies and the drainage system to create resilience cities against future floods and protect humans, assets, infrastructure, and properties from damages (Bertilsson et al., 2019; Carter et al., 2015; Vercruysse et al., 2019). Nowadays, drainage systems in cities are under pressure due to urbanisation and are not able to withstand higher intensity and frequency of storm events (Eulogi et al., 2021; Rosenzweig et al., 2018). As a result, flood risk management adaptation solutions are currently blocked by the lack of affordable and feasible strategies. In some studies of adaptation strategies at local scales with Sustainable Urban Drainage Systems (SuDS) (Fletcher et al., 2015) and larger scales (Li et al., 2018) there is a non-adaptation efficiency apparent between them (Ernst & Preston, 2017; Kuller et al., 2017; Preston et al., 2015) due to the lack of information and uncertainty of the capacity, the performance and the location of these systems (Hoang & Fenner, 2016; Mailhot & Duchesne, 2010; O'Donnell et al., 2017; Schuch et al., 2017). A recent study by Oladunjoye (2022) asserts that SuDS could be a crucial component of flood mitigation due to the capacity to decrease runoff volume and lower the danger of floods by controlling the flow in natural infiltration systems.

While the potential of more realistic and affordable solutions such as Blue-Green Infrastructure (BGI) or Natural Flood Management (NFM) is increasingly recognised, the barriers to their uptake are formidable, as effective city-wide schemes require significant investment in implementation in multiple locations if they are to be effective, rather than opportunist and piecemeal schemes where re-development permits. A more systematic city-wide approach is therefore required, with a clear demonstration of the overall cost and benefits, before a city planning authority will be prepared to invest. Such an approach requires an urban flow simulation with sufficient detail to resolve not only the exposure (and subsequent benefit form risk reduction) at property level, but also the highly localised runoff sources and flow paths where BGI can be implemented. This study therefore sets out to advance a new and systematic methodology combining an advanced high-resolution hydrodynamic flood model with a source-receptor benefit cost methodology.

Urban flood models have been developed over the last decade (Glenis et al., 2018; Guo et al., 2021; Sanders, 2017; Teng et al., 2017; TUFLOW, 2018) to better understand the flood dynamics, better estimate the water flow paths and the flood depths around cities, and can increasingly be used to investigate the connectivity of flood management options and interventions with the characteristics of a city (impermeable surfaces, topography, storms etc). If applied at high resolution and large enough scales, these hydrodynamic models can provide accurate analysis of future flood risk, and with the collaboration of local authorities can be used to design flood mitigation solutions by locating areas at high flood risk and adding interventions in critical areas to reduce pluvial or fluvial flooding (Alves et al., 2016; Casal-Campos et al., 2015; Dawson et al., 2020; Hewett et al., 2020; Kapetas & Fenner, 2020; McKenna et al., 2023; Morgan & Fenner, 2019; Singh et al., 2021). This study demonstrates the use of such an advanced high resolution system, the CityCAT hydrodynamic model (Glenis et al., 2018), in a systematic framework to locate optimal areas where interventions can be made, accounting not only for their cost, but also their benefit in reduction of damages to properties from flooding in a city scale catchment. Recent studies have begun to explore methods of optimising location of BGI interventions, such as Birkinshaw and Krivtsov (2022) who established that for a particular urban setting locating a retention pond further upstream was most effective, and that storage in the lower part of the catchment could actually increase flood risk. There is however a need to more systematically manage the assessment of such options, so a cost-benefit driven *'source-receptor'* analysis is developed here to locate areas at high flood risk, how many buildings are impacted by flooding, what information we extract from the model and where critical interventions should be added to the model to most efficiently reduce cost and flood damages.

## 2. Methodology

### 2.1. Case study

The novelty of the cost-benefit driven *'source-receptor'* flood risk framework offers the flexibility for designs to be developed in every urban area or catchment with any suitable resolution of the Digital Terrain Model (DTM) (according to Christos Iliadis et al. (2023a), a DTM resolution of less than 5m is required to capture all the required information for this framework) to any commercial or research flood model. The required information and inputs of the methodology are: a) the Digital Elevation Model or Digital Terrain Model (DEM/DTM) of the study area; b) the buildings (classification, e.g. commercial or residential) and green spaces; c) rainfall data - the construction of IDF/DDF curves or storm profiles are needed to specify a range of flood hazards and d) flood damage estimates for commercial and residential buildings of the study area. In this study, the campus of Newcastle University and the adjacent city centre are subject to a major flood risk from upstream and suffered major damages during the 2012 *'Toon Monsoon'* thunderstorm event (Kutija et al., 2014). The area is characterised by historic and commercial buildings, residential properties, and green space with the most important being parkland areas, primarily the Town Moor and Leazes Park, which cover a significant extent of the study area. Previous studies have been made in relation to pluvial flooding after the historic storm, on the 28th of June 2012, to add BGI in critical places in order to protect the assets (Fenner et al., 2019; Kilsby et al., 2020; O'Donnell et al., 2020; Wright & Thorne, 2014).

The catchment has been modelled using CityCAT (Glenis et al., 2018) for storm events spanning 1 in 10-year to 1 in 100-year return period with a duration of 60 min. Figure 1 shows the study catchment with buildings colour coded according to use, and land use. The resolution of the DEM and computational grid is two metres (area of each grid square is 4m$^2$) and was derived from Lidar (2016). The total number of computational cells in the domain is 1,005,904 covering an area of 5.30 km$^2$. The numerical grid was generated following the *'Building Hole'* approach where building footprints are removed from the computational domain by generating a non-flow boundary around them, which is more realistic than widely used approximations such as the so-called stubby-building approach. Rainfall on to the building is retained by re-distribution to the nearest surface grid square (for a full description, see Christos Iliadis et al. (2023b)).

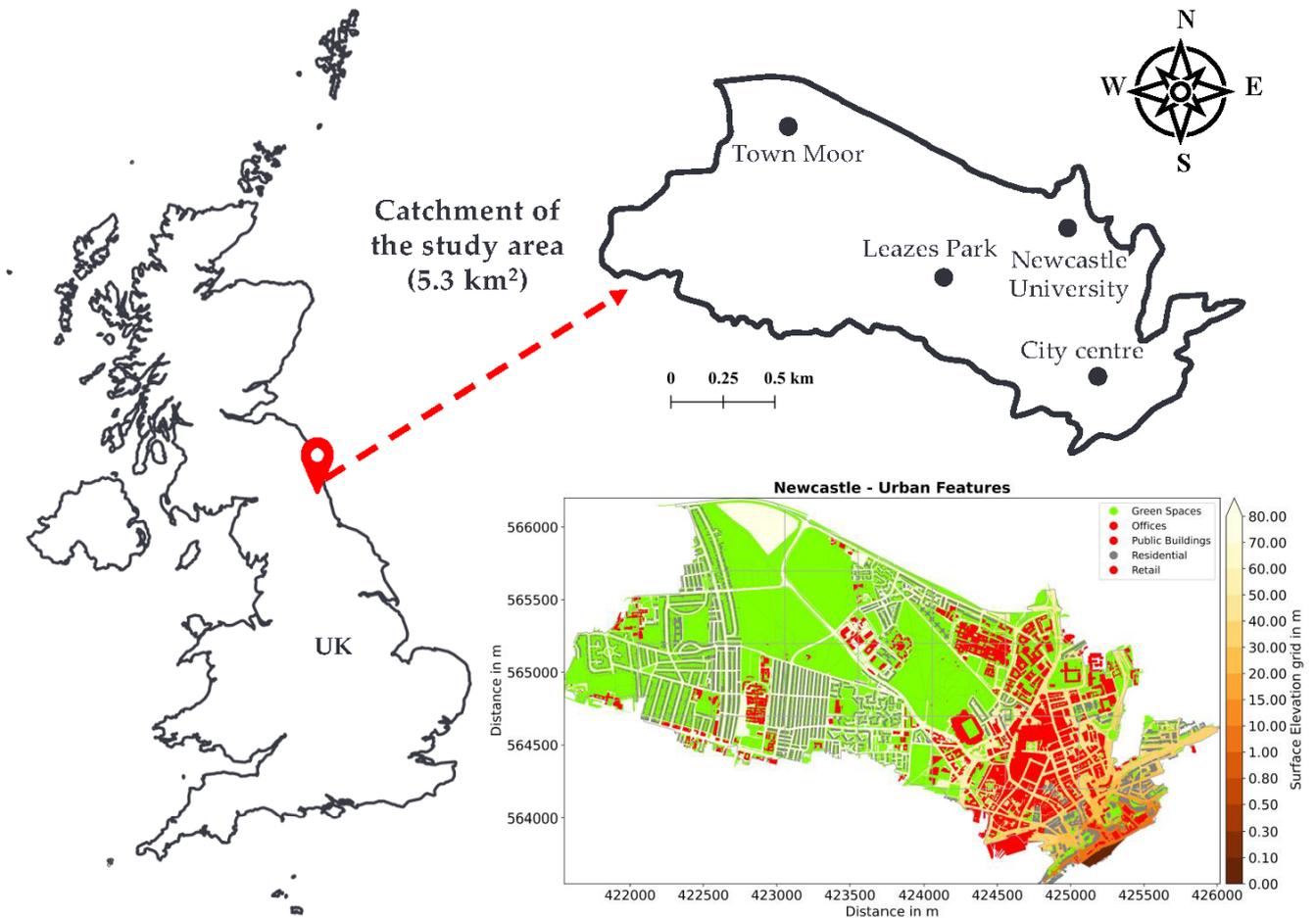

**Figure 1:** Overview of the study area in Newcastle upon Tyne, UK. Grey represents residential buildings, red commercial (offices, public buildings, retail), green is permeable areas, and brown to yellow shading the ground surface elevation.

*2.2. A cost-benefit 'source-receptor' flood risk framework*

A powerful way to identify areas contributing to the total flood extent during a simulated event is the *'source-to-impact'* flood analysis based on a systematic cell dependency applied by Vercruysse et al. (2019) and Dawson et al. (2020) for the urban core of Newcastle upon Tyne and by Ewen et al. (2013) for the river Hodder catchment in the northwest England. The analysis of Vercruysse et al. (2019) presented a four step methodology to capture the flow paths and identify the location at high flood risk. They considered the differences of modelled maximum water depths generated by subtracting the baseline scenario with the cells without rainfall of the catchment, and four spatial prioritization criteria to identify the best cells in which to add interventions. In many cases, maximum values are not suitable for this purpose due to the instabilities in the accuracy of elevation (*'errors'* in the DEM/DTM) which can result in the overestimation of flood values in places with minor hazard (Huang et al., 2022; Christos Iliadis et al., 2023b; McClean et al., 2020).

The framework developed in this study combines extreme rainfall information, flood dynamics and the cost-benefits of flood risk management in an urban area. The methodology consists of five steps to (i) identify the water flow paths; (ii) capture the rainfall for a range of different magnitude storm events; (iii) categorise buildings at significant flood risk; (iv) calculate the damages from flooding; and (v) add interventions in critical high-risk locations upstream or downstream prioritised according to their cost-benefit. The steps of the cost-benefit *'source-receptor'* framework are detailed below:

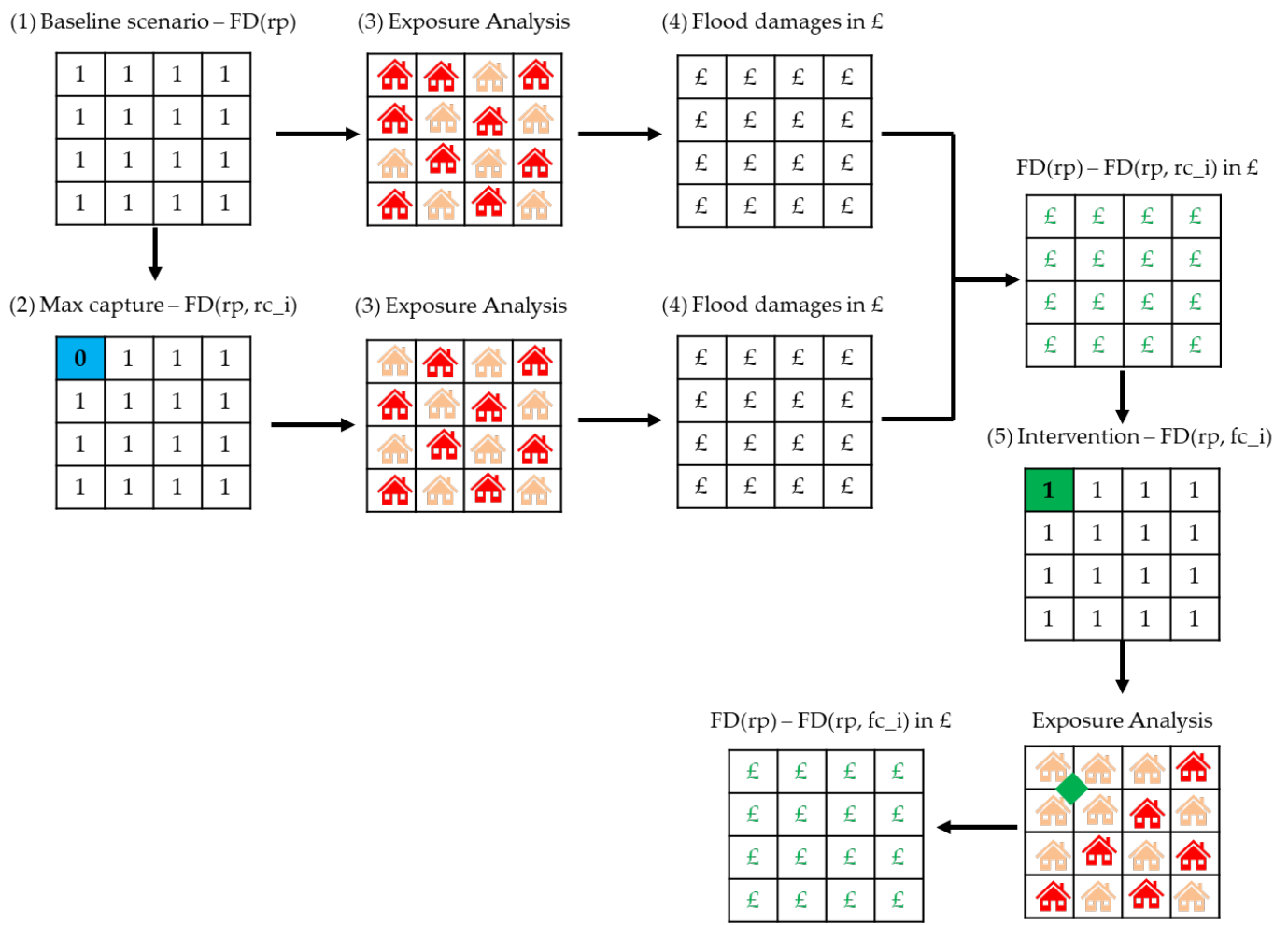

**Figure 2:** Schematic workflow of the cost-benefit *'source-receptor'* flood risk methodology.

1. *Divide the study area into approximately equal size cells and classify the type of buildings;*

2. *Run the CityCAT model to generate flood depths for equal rainfall across all cells for the four different storms – Baseline Scenario FD(rp), where FD is the flood damages and rp is the return period of the storm;*

3. *Turn off rainfall in individual cells – we refer to this as Max Capture FD(rp, rc_i), where rc_i is a rainfall cell and i = 1..Total number of rainfall cells;*

4. *Classify the buildings according to their flood risk (exposure analysis) and calculate the flood damages for the baseline scenarios and the max capture scenarios;*

5. *Calculate the benefit by subtracting the max capture scenarios damages from the baseline scenarios flood damages FD(rp) – FD(rp, rc_i);*

6. *Add interventions to cells (such as permeable pavements, water butts, green roofs, storage ponds etc) FD(rp, fc_i), where fc_i is an intervention (or "feature") cell and i = 1..Total number of intervention cells;*

7. *Classify the buildings according to their flood risk (exposure analysis) and calculate the damages;*

8. *Compare the baseline scenarios with the intervention scenarios to identify the best cost-benefit solution to reduce flooding FD(rp) – FD(rp, fc_i);*

The outcomes of the analysis are of course dependent on the size of the cells within the grid, the spatial resolution of the DEM/DTM, the type of buildings and the available green space in the study area. The first step is to divide the catchment into equal cells, twenty-three in this study, with an area of 500m × 500m, to evaluate the influence of spatial delineation on source areas, a sensitivity analysis could be undertaken by varying the cell size according to the study area, approximately which would be considered as *'source area'* for the surface runoff and classify the type of buildings (commercial and residential) in the study area. The second step is to run the CityCAT model to generate the baseline scenario (*FD(rp)*) for multiple storm events: here four different storm magnitudes were used, which cover the range of storms required to estimate the flood exposure of the buildings, and the damages from flooding in the study area. We use storms corresponding to 1 in 10-year, 1 in 20-year, 1 in 50-year and 1 in 100-year (similar to the historic storm *'Toon Monsoon'*) return period with a duration of 60 minutes. The third step is to one-by-one switch off the rainfall in every cell of the study area (*FD(rp, rc_i)*) and then run the CityCAT model multiple times (i.e. one run per cell per storm scenario, in this study a total of 92 runs). This represents a total *'capture'* of the rainfall which means that there is no runoff from that cell. Moreover, the fourth step is to estimate the flood exposure to buildings and the flood damages per max capture scenario. In addition, the next step is to compare the baseline flood damages with the max capture damages by subtracting both for every cell (*FD(rp) – FD(rp, rc_i)*). Then, this cost-benefit step and the available green space in every cell allows ranking the areas from high-priority to low-priority cells to add adaptation solutions to mitigate runoff.

A range of interventions such as permeable pavements, water butts, green roofs, and storage ponds (SuDS) can be explicitly represented in CityCAT. Hence, the next step is to locate the areas at high flood risk through the provided information from steps 1 to 5 and the connectivity between the damages, the available green space and flood source areas where interventions can be implemented (see section 3.3.) in order to add interventions to these cells/areas (*FD(rp, fc_i)*), and run again the model equal times as the added interventions for different storms, in this study permeable pavements and ponds in critical locations were used. Then, the new flood exposure to buildings (Bertsch et al., 2022) and the damages are calculated to check the behaviour of interventions against multiple storms. Finally, the baseline scenarios are compared with the intervention scenarios (*FD(rp) – FD(rp, fc_i)*) to identify the best cost-benefit solution to reduce flood damages, and explore if the proposed adaptation options for their building are acceptable to local authorities and insurance companies to reduce damages, content the properties and increase their resilience against the direct contact of flood water (Priest et al., 2022).

*2.3. Assessing flood risk – flood exposure to buildings caused by each grid square*

The most important criterion of the cost-benefit *'source-receptor'* flood risk framework to locate areas at high flood risk is a novel flood exposure analysis (Bertsch et al., 2022) to estimate the flood risk likelihood to buildings by analysing the water depths adjacent to the building (i.e. a one grid square buffer). These depths could be analysed as the mean depth in a buffer zone around the building, or the maximum depth, or more robustly the $90^{th}$ percentile depth (used in this study) to avoid undue influence of a single erroneous depth value. The buffer zone depends on the computational grid resolution, 2m in this study. If the depths exceed a threshold of 30 cm (see Table 1) then buildings can be classified as high, medium, and low risk. In addition, the buildings at low flood risk have been excluded here by assuming that the damages from flooding are minor in comparison to damages to buildings at high and medium risk.

**Table 1:** The criteria for calculating flood exposure likelihood for buildings.

| Exposure Class | Mean depth (m) | 90$^{th}$ percentile (m) |
|---|---|---|
| Low | <0.10 | <0.30 |
| Medium | <0.10 | ≥0.30 |
|  | ≥0.10 - <0.30 | <0.30 |
| High | ≥0.10 | ≥0.30 |

*2.4. Flood damages to properties*

In the UK, flooding causes average damages of £1.3 billion per year, the cost for flood defence is around £4.4 billion the last decade, and the properties at flood risk are more than 5.2 million in England alone (Craig, 2021; Environment Agency, 2022; UK Government, 2016). Hence, residential and commercial flood damage is a crucial case that the researchers, the local authorities, and insurance companies collaborate with each other to propose efficient and innovative solutions against flooding. The available buildings in the study area are commercial (retail, public buildings, offices) and residential. The cost-benefit *'source-receptor'* flood risk framework took values (corresponding to 2022 prices) to calculate the damages to commercial and residential buildings from the Handbook for Economic Appraisal (Multi-Coloured Handbook, 2022) by Priest et al. (2022) (Figure 3). Note that prices may differ for other parts of the UK and definitely for other countries.

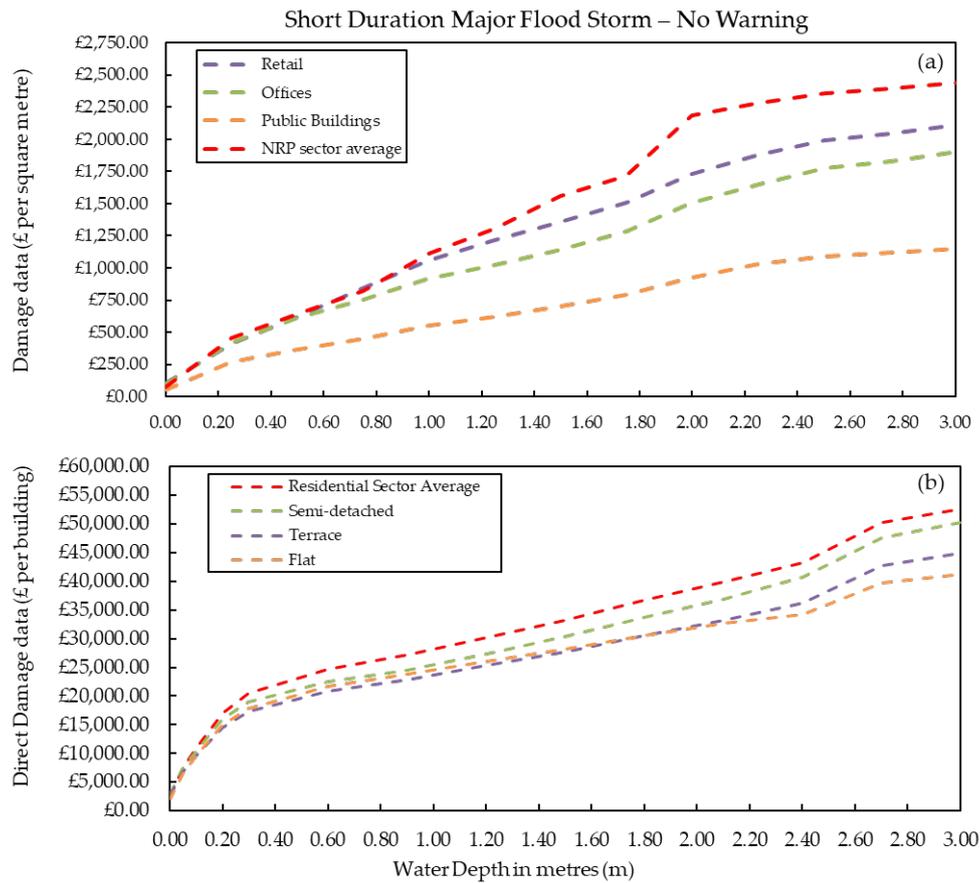

**Figure 3:** Direct damage from different water depths for: (a) commercial buildings; (b) residential buildings (Priest et al., 2022).

*2.5. Land in green spaces in cells*

The only spatial criterion in this framework covers the percentage of land use in the study area (Figure 4) where flooded green spaces may be considered as the areas most suitable to add efficient interventions such as ponds or swales to protect assets downstream or to guide the researcher to add other types of interventions such as permeable pavements, water butts etc. to other parts of the catchment.

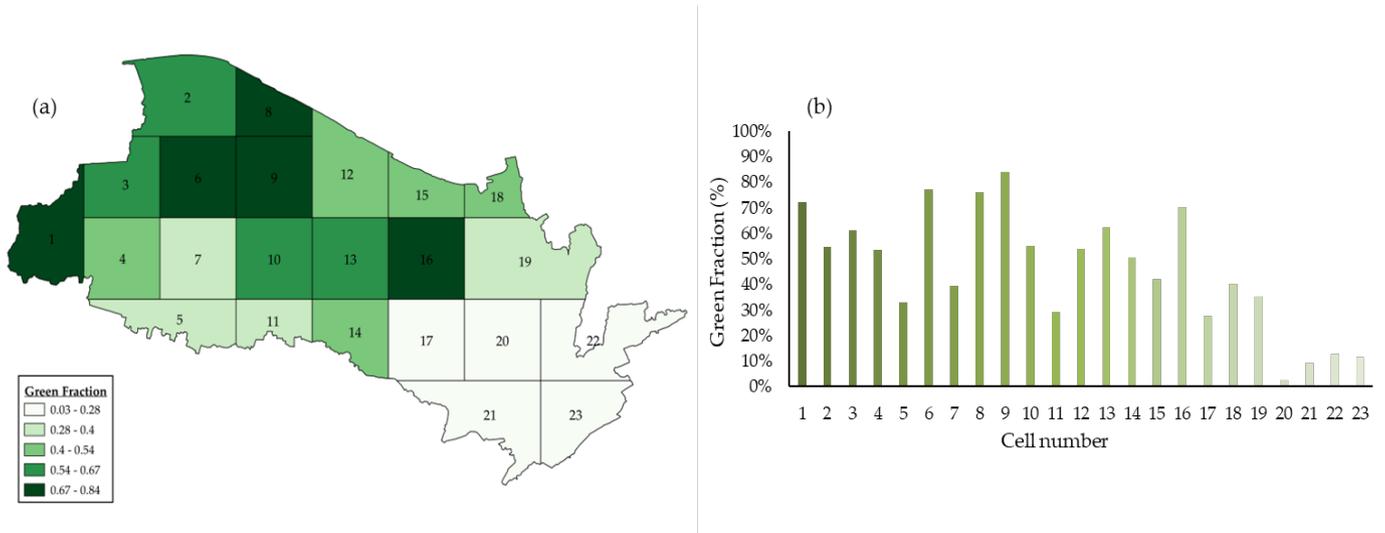

**Figure 4:** The percentage of green spaces on the cells of the study area: (a) model green spaces; (b) the summary statistics.

## 3. Results

### 3.1. Baseline flood results

The modelled flood depths for the baseline scenarios (*FD(rp)*) shows us that in the catchment there are two main flow paths, the first from the west side of the catchment to the east (cells 1, 4, 6, 7, 9, 12, 15, 18 and 19) through Newcastle University campus and the second through the city centre in the lower catchment (cells 6, 9, 10, 13, 14, 17, 20 and 21), see Figure 5. The flood exposure to buildings was calculated for the baseline scenarios to identify the number of buildings at high and medium flood risk and the cells that caused flooding to them (Table 2, Figure 5). Most of the buildings at high flood risk are located in the west, central, and downstream of the catchment, which is to be expected due to the generated flow paths in cells of the study area and the different characteristics of the ground, e.g. impermeable pavements, in the study area.

**Table 2:** Number of inundated buildings per baseline scenario for different storm events.

| FD(rp)   | Medium | High | Total |
|----------|--------|------|-------|
| FD(10y)  | 206    | 258  | 464   |
| FD(20y)  | 272    | 396  | 668   |
| FD(50y)  | 411    | 627  | 1038  |
| FD(100y) | 518    | 809  | 1327  |

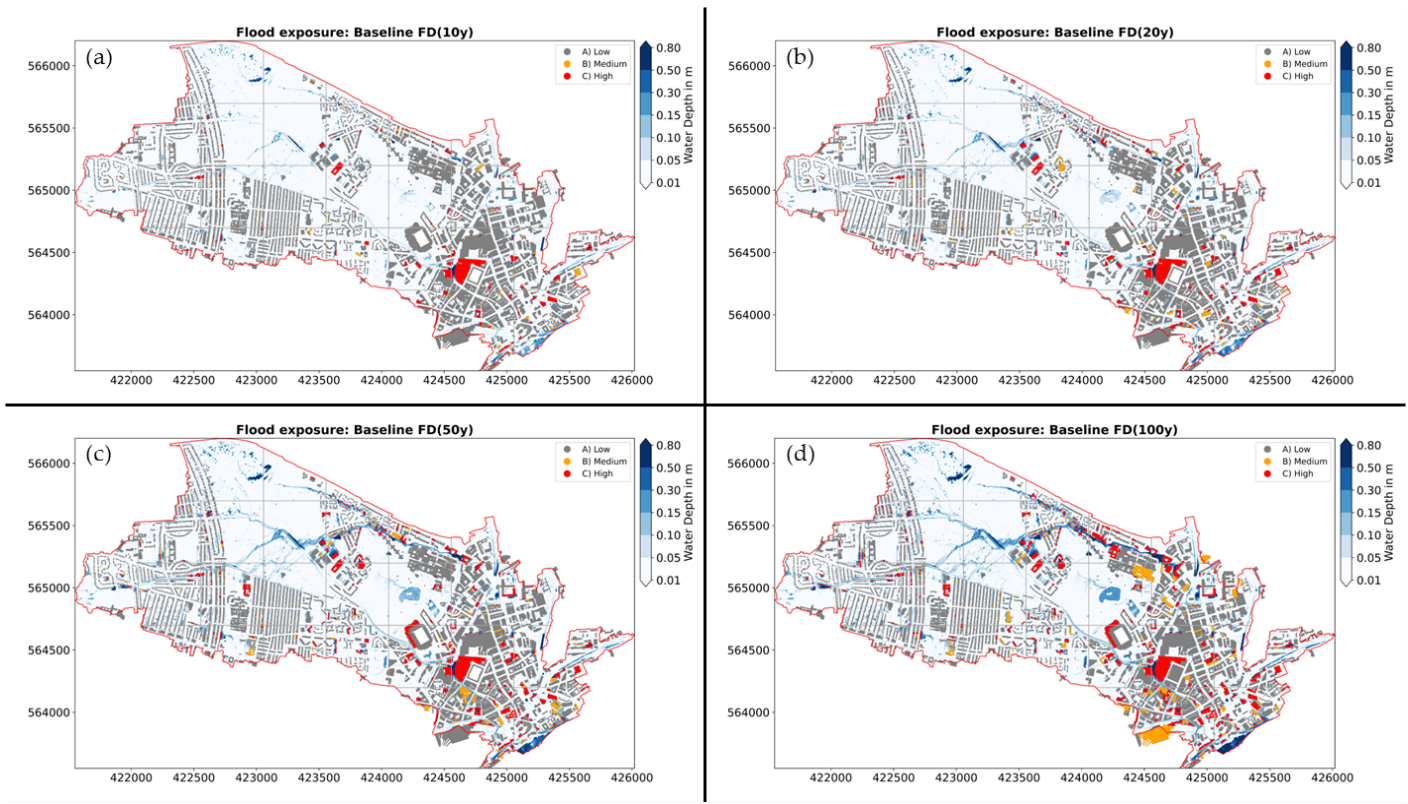

**Figure 5:** Flood depth from CityCAT simulation and flood exposure to buildings for the baseline scenarios - *FD(rp)* for: (a) a 1 in 10-year storm event; (b) a 1 in 20-year storm event; (c) a 1 in 50-year storm event; and (d) a 1 in 100-year storm event with a duration of 60 min, the red colour defines the building at high risk, the orange at medium risk and the grey at low risk for Newcastle city centre.

The estimated total flood damages per baseline scenario can be seen in Table 3 and in Figure 6, where the total damages even for the *'small'* storm events (1 in 10-year and 1 in 20-year return period) are high. This is consistent with the significant commercial buildings which are impacted in the centre of the catchment. Note here that from the exposure analysis for the baseline scenarios some buildings are classified at high and medium flood risk but after turning off the rainfall to cells the classification scheme changes to some buildings from high to medium and low further downstream as expected (see Figure 7: further flood maps are available as supplementary material).

**Table 3:** The total flood damages for the baseline scenarios – *FD(rp)*.

| FD(rp)   | Commercial (£) | Residential (£) | Total Damages (£) |
|----------|----------------|-----------------|-------------------|
| FD(10y)  | £40.8M         | £6.1M           | £47.0M            |
| FD(20y)  | £59.5M         | £8.9M           | £68.5M            |
| FD(50y)  | £92.3M         | £14.3M          | £106.7M           |
| FD(100y) | £147.9M        | £18.4M          | £166.5M           |

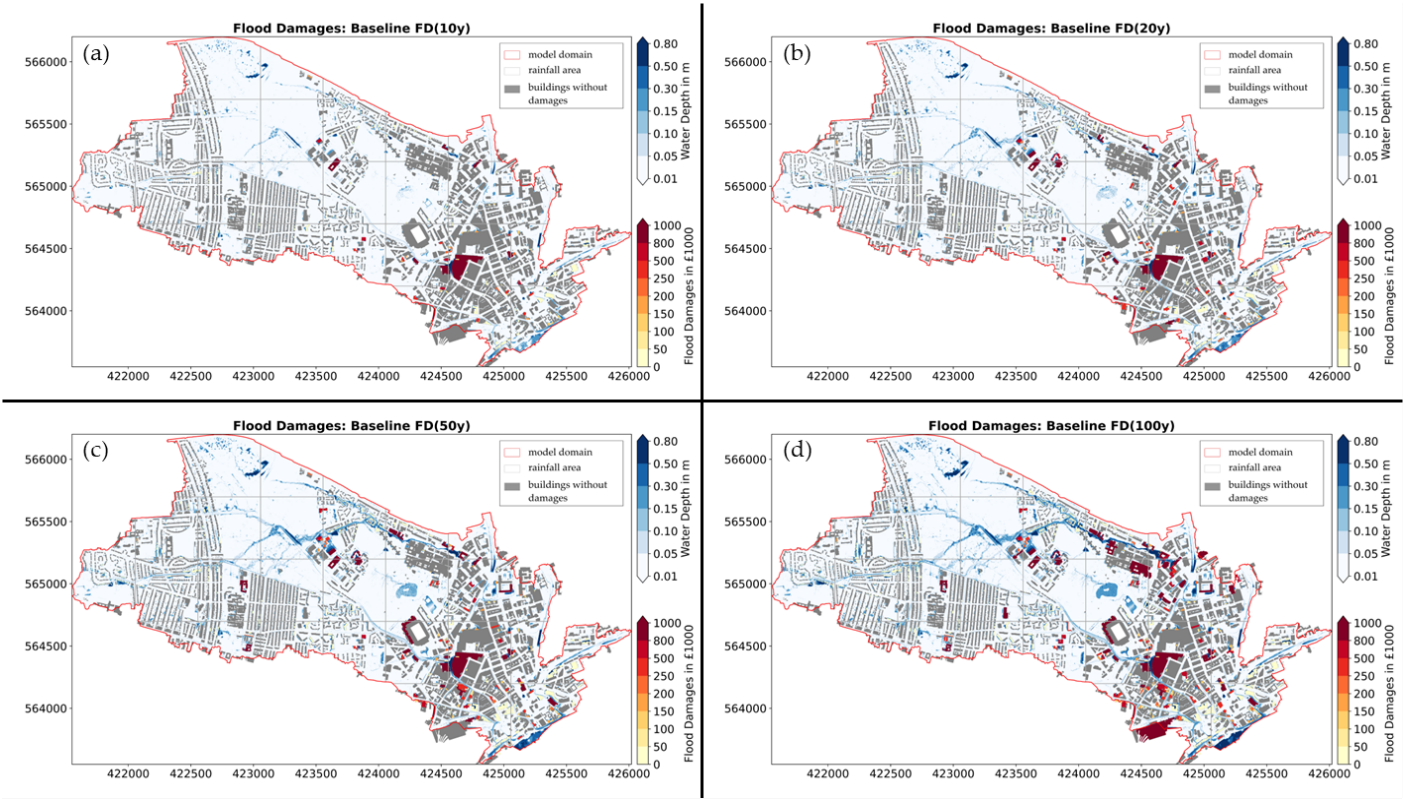

**Figure 6:** Examples of flood damages and water depth maps for the baseline scenarios – *FD(rp)* for: (a) a 1 in 10-year storm event; (b) 1 in 20-year storm event; (c) 1 in 50-year storm event; and (d) 1 in 100-year storm event with a duration of 60 min, yellow to dark red defines the cost per buildings from flooding.

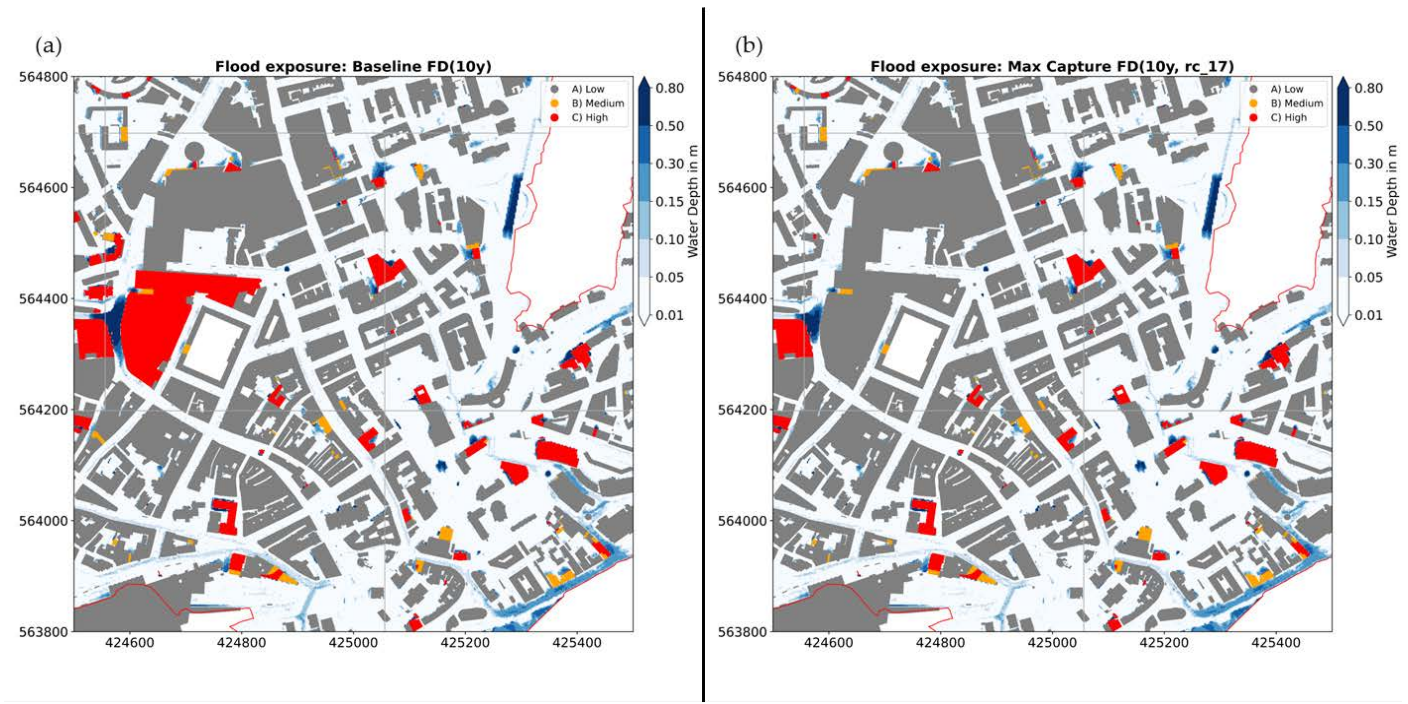

**Figure 7:** Example of flood exposure map, before (left) and after (right) turning off rainfall in cell 17 *(FD(10y, rc_17))*, the red colour defines the buildings at high risk, orange at medium risk and the grey at low risk.

*3.2. Rainfall capture*

The cost-benefit *'source-receptor'* flood risk framework was developed to assess the impact of certain cells on surface flooding by analysing the exposure to buildings and calculating the flood damages to properties on a local/large scale and further downstream. The simulated flood depths from the baseline scenarios (*FD(rp)*) for the multiple storm events allow us to identify the flow paths in the study area. Next the exposure analysis allows us to locate the buildings (commercial & residential) at high/medium flood risk and then the difference in flood damages to buildings from (a) the baseline scenarios (*FD(rp)*) and (b) the max capture scenarios (*FD(rp, rc_i)*) represents the benefit (damage reduction) to buildings by switching-off the rainfall to cells (i.e. capturing all the rainfall in the cell – the ideal maximum intervention). The matching of cells identified in this way with the available green space offers the capability to identify high-priority cells to add adaptation solutions (Table 4, Figure 8 and 9) in a straightforward way. The highest value in the final column (Cost-Benefit * Green Fraction (GF)) of Table 4 corresponds to the highest priority location and this is used to select the location and type of intervention most suitable to build: this is discussed further in section 3.3. Moreover, due to simulating for multiple storm scenarios, it can be seen in Figure 9 that the prioritisation of cells varies for different magnitudes of storm (*FD(rp, rc_i)*) which is to be expected due to the different flow paths in the catchment and the different extent of rainfall capture in cells.

**Table 4:** The ranking system to prioritise cells with a high need of intervention for the different storm events, the five high-priority cells for all the max capture scenario per storm event, damages in £ million.

| Cells | Green Fraction (GF - %) | Commercial (£) | Residential (£) | Total Damages (£) | Cost-Benefit (FD(10y) – FD(10y, rc_i)) in £ | Cost-Benefit * GF |
|---|---|---|---|---|---|---|
| **10-year return period** | | | | | | |
| FD(10y, rc_13) | 62.20% | £35.7M | £5.8M | £41.5M | £5.5M | 3.38 |
| FD(10y, rc_17) | 27.71% | £29.2M | £6.1M | £35.3M | £11.7M | 3.24 |
| FD(10y, rc_14) | 50.70% | £36.2M | £6.1M | £42.3M | £4.7M | 2.37 |
| **20-year return period** | | | | | | |
| Cells | Green Fraction (GF - %) | Commercial (£) | Residential (£) | Total Damages (£) | Cost-Benefit (FD(20y) – FD(20y, rc_i)) in £ | Cost-Benefit * GF |
| FD(20y, rc_13) | 62.20% | £52.1M | £8.6M | £60.9M | £7.6M | 4.79 |
| FD(20y, rc_17) | 27.71% | £48.4M | £8.9M | £57.3M | £11.2M | 3.08 |
| FD(20y, rc_12) | 53.82% | £55.7M | £8.2M | £63.9M | £4.6M | 2.47 |
| **50-year return period** | | | | | | |
| Cells | Green Fraction (GF - %) | Commercial (£) | Residential (£) | Total Damages (£) | Cost-Benefit (FD(50y) – FD(50y, rc_i)) in £ | Cost-Benefit * GF |
| FD(50y, rc_13) | 62.20% | £82.2M | £14.0M | £96.2M | £10.5M | 6.53 |
| FD(50y, rc_16) | 70.29% | £86.3M | £13.8M | £100.1M | £6.6M | 4.64 |
| FD(50y, rc_17) | 27.71% | £76.4M | £14.2M | £90.6M | £16.1M | 4.45 |

| | | | 100-year return period | | | |
|---|---|---|---|---|---|---|
| Cells | Green Fraction (GF - %) | Commercial (£) | Residential (£) | Total Damages (£) | Cost-Benefit (FD(100y) – FD(100y, rc_i)) in £ | Cost-Benefit * GF |
| FD(100y, rc_16) | 70.29% | £135.0M | £17.8M | £152.8M | £13.7M | 9.56 |
| FD(100y, rc_9) | 84.13% | £138.9M | £18.1M | £157.0M | £9.5M | 7.98 |
| FD(100y rc_13) | 62.20% | £137.9M | £18.1M | £156.0M | £10.5M | 6.50 |

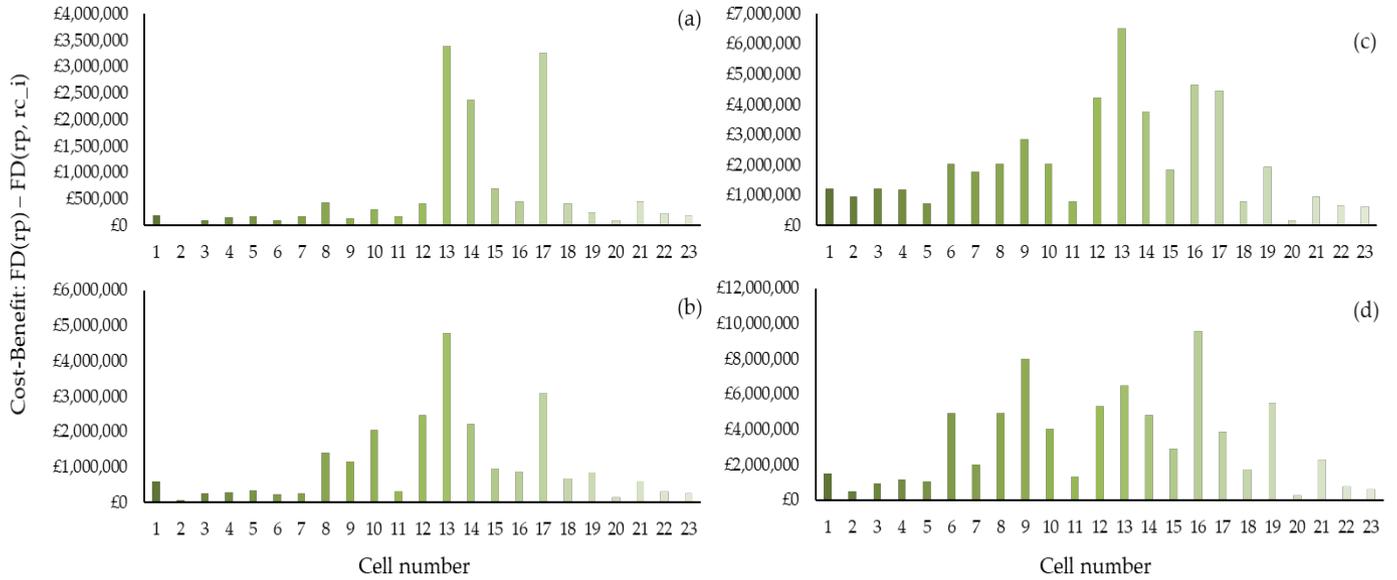

**Figure 8:** Summary statistics of the cost-benefit *'source-receptor'* for the max capture scenarios for all the storm events: (a) 1 in 10-year return period; (b) 1 in 20-year return period; (c) 1 in 50-year return period; and (d) 1 in 100-year return period with a duration of 60 min.

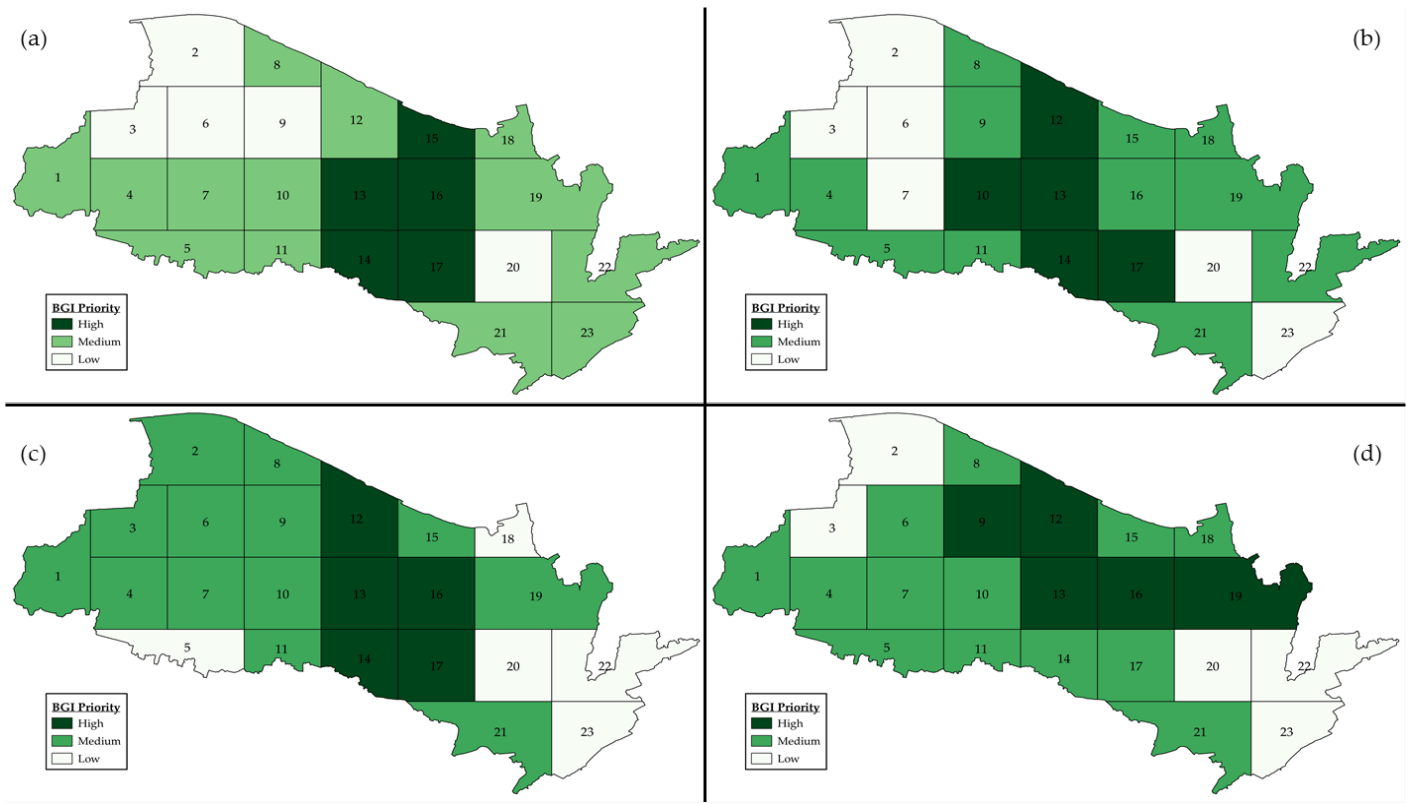

**Figure 9:** Classification of cells as priority areas for BGI in the study area: (a) 1 in 10-year return period; (b) 1 in 20-year return period; (c) 1 in 50-year return period; and (d) 1 in 100-year return period with a duration of 60 min.

*3.3. Adding Blue-Green Infrastructure to urban areas and catchments*

The previous section has examined the use of BGI to capture rainfall directly on receipt at the ground, but the cost-benefit *'source-receptor'* flood risk framework can equally well be used with BGI interventions to capture runoff, for example with permeable pavements or detention ponds.

*3.3.1. Runoff Capture – permeable pavements*

Firstly permeable pavements are introduced to capture runoff for storm events at 1 in 10-year and 1 in 20-year return period with a duration of 60 min. The results from the exposure, the cost-benefit analysis of buildings and the ranking system for the small events (see supplementary material for the tables and the flood maps) suggest location of adaptation in cells 13, 17, 14, 15 and 16 for a 1 in 10-year storm event and to cells 13, 17, 12, 14 and 10 for a 1 in 20-year storm event. Following identification, permeable pavements (*FD(rp, fc_i)*) were introduced in these cells to estimate the benefit reducing damages to buildings downstream. The installation cost for permeable pavements is around £30 per square metre of pavement according to SNIFFER (2006) and Woods Ballard et al. (2015) with an operational cost of £0.40 per square metre per year (Gordon-Walker et al., 2007) and 40 years of paving life. Table 5 below describes the cost-benefit in cells classified as high priority with the area of pavements in these cells and the installation cost to identify the most economic cell to add the proposed BGI. It can be seen that for cell 17 the reduction in damages is almost £1.60M by adding permeable pavements with a cost of £0.65M (Figure 10).

**Table 5:** The flood damages for the intervention scenarios, add permeable pavements in cells 13, 14, 15, 16 & 17, the cost-benefit, and the installation cost.

| | **Intervention Scenarios 10-year return period** | | | | | |
|---|---|---|---|---|---|---|
| **Cells** | **Commercial (£)** | **Residential (£)** | **Total Damages (£)** | **Cost-Benefit (FD(10y) – FD(10y, fc_i)) in £** | **Area (m$^2$)** | **Installation cost (£)** |
| FD(10y, fc_13) | £40.7M | £6.1M | £46.8M | £0.24M | 15,396.718 | £0.46M |
| FD(10y, fc_14) | £40.3M | £6.1M | £46.4M | £0.64M | 21,468.389 | £0.64M |
| FD(10y, fc_17) | £39.3M | £6.1M | £45.4M | £1.64M | 21,726.375 | £0.65M |

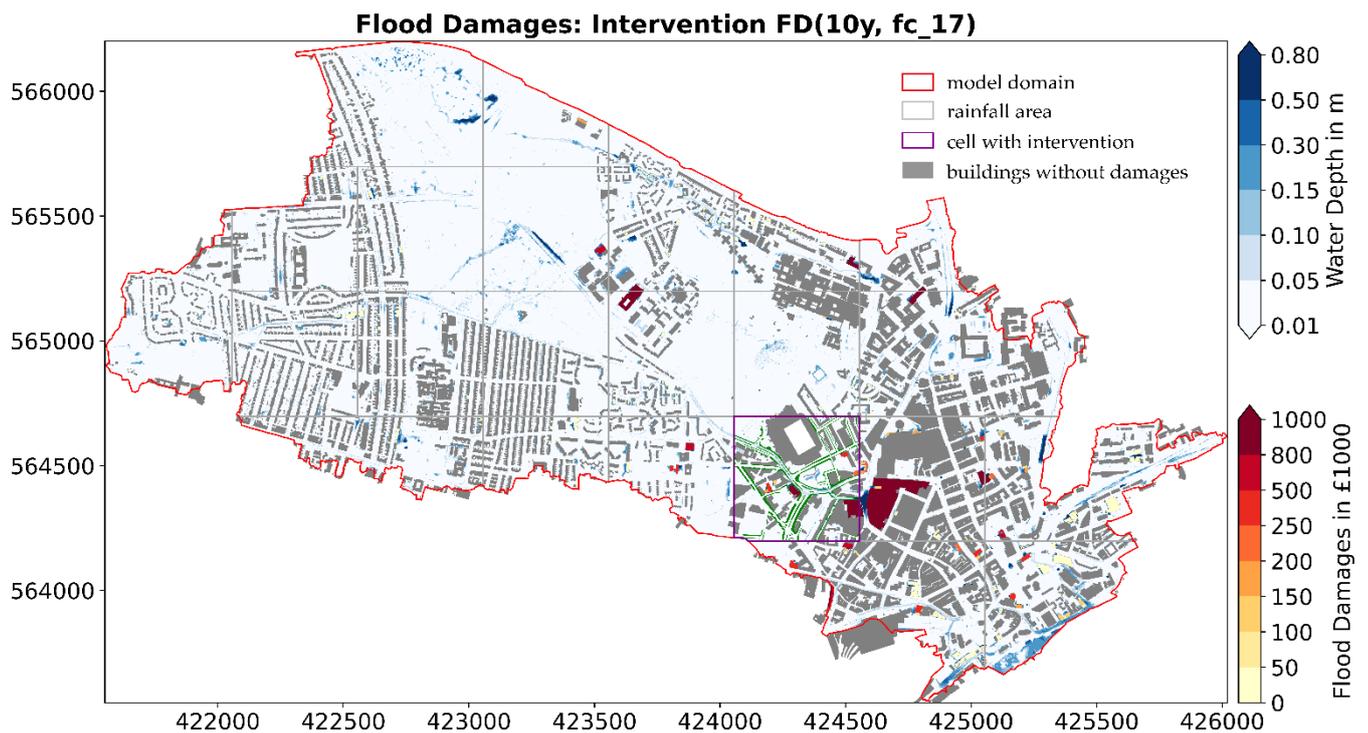

**Figure 10:** Example of flood damages and water depth map, add an intervention in cell 17, green denotes the permeable pavements, yellow to dark red defines the cost per buildings from flooding.

### 3.3.2. Runoff Capture – detention ponds

A second method of runoff capture, installing detention ponds, has been examined targeted at expendable green spaces. This builds on previous work (Birkinshaw & Krivtsov, 2022) which assessed the optimal location for ponds. Two cells (9 and 12, see Figure 11) were chosen for further investigation as they are classified as high-priority cells with a high percentage of available green space (84% and 54% respectively) and a detention pond (*FD(rp, fc_i)*) was proposed to be built in each cell. The first step is to choose manually the best place in these cells to add the pond and re-run the model to estimate the reduction in flood damages. In cell 12 the total area of the proposed pond is 510 m$^2$ with corresponding volume of 765 m$^3$ and in cell 9 the area of the detention pond is 8,000 m$^2$ with a volume of 10,000 m$^3$. The estimated flood damages, the cost-benefit, and the installation cost by simulating the ponds in cells 9 and 12 for multiple storm events can be seen in Table 6. An average cost to construct a pond is £80,000 per 5,000 m$^3$ water volume, so the estimated cost to build the two detention ponds in cells 9 and 12 is around £0.16M with the economic benefit being more than £8.5M for the high storm events and more than £0.50M for small events (SNIFFER, 2006; Woods Ballard et al., 2015). Thus, the pond life could be more than 15 years with an efficient maintenance cost of £0.60 per square metre per year (Gordon-Walker et al., 2007). Finally, a combination of interventions could be proposed to model in areas where the intensity of rainfall is extremely high (e.g. > 100 mm of rainfall).

**Table 6:** The flood damage costs for the intervention scenarios, add storage pond in cells 9 and 12 for multiple storm events, the cost-benefit, and the installation cost, see Table 3 for the baseline scenario damages *FD(rp)*.

| Cells | Commercial (£) | Residential (£) | Total Damages (£) | Cost-Benefit (FD(rp) – FD(rp, fc_i)) in £ | Installation cost (£) |
|---|---|---|---|---|---|
| FD(20y, fc_9_12) | £59.1M | £8.9M | £68.0M | £0.55M | ≈ £0.16M |
| FD(50y, fc_9_12) | £90.3M | £13.7M | £104.0M | £2.75M | ≈ £0.16M |
| FD(100y, fc_9_12) | £140.0M | £17.7M | £157.7M | £8.78M | ≈ £0.16M |

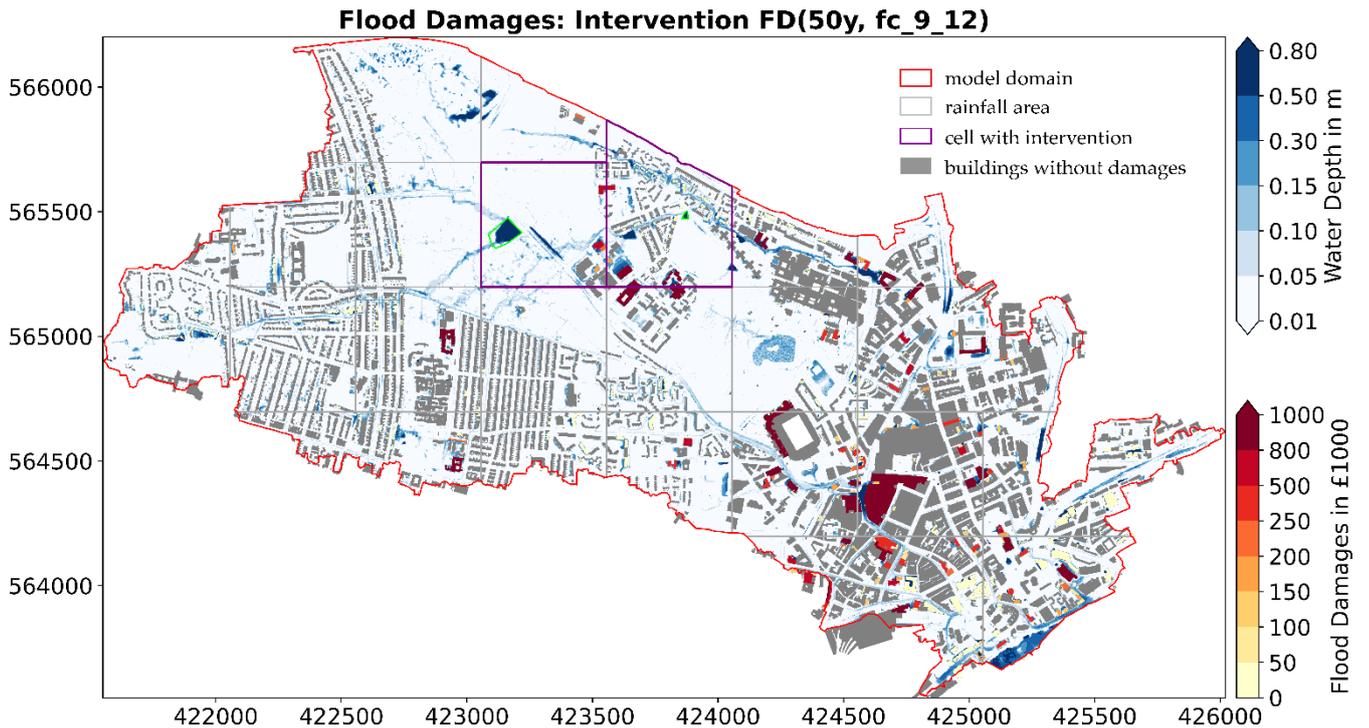

**Figure 11:** Example of flood damages and water depth map, with intervention in cells 9 and 12 (*FD(50y, fc_9_12)*, storage ponds), green denotes the ponds, yellow to dark red defines the damages per buildings from flooding.

## 4. Discussion

To assess the cost-benefit using the '*source-receptor*' analysis suggested in this study, it is important to evaluate the significance of damage reduction in terms of benefits. The proposed permeable pavements in cell 17 during a storm event with low intensity (10 years return period) could yield a benefit of almost 4% of the total damages in the case study. Additionally, the suggested ponds in cells 9 and 12 could contribute to a benefit of up to 6%. However, the proposed framework may demonstrate greater efficiency in other urban areas where a combination of interventions, such as water butts, green roofs, etc, is implemented. The study area of this research is characterised by high instabilities in surface elevation and steep slopes, only the proposed interventions (permeable pavements and ponds) prove to be more effective compared to other adaptation solutions, which will be studied further in future research.

It is crucial to recognise the constraints that define the scope of the outcomes. A framework such as the cost-benefit '*source-receptor*' often encounters practical challenges in real-world complexities, such as fragmented land and property ownership, which could present significant obstacles to coordinating and implementing comprehensive adaptation solutions across diverse stakeholders. Achieving a balance between the theoretical effectiveness of risk reduction measures and their practical feasibility requires addressing these intricate challenges, streamlining procedures, and fostering collaborative efforts among stakeholders, insurance companies, and local authorities to ensure long-term success of risk mitigation strategies.

This study has prioritised economic considerations, but it must be recognised that other aspects may be equally important. Restricting aspects include not only land ownership as outlined above, but also acceptance by communities and stakeholders on aesthetic, access or safety grounds. Positive aspects to increase benefit are increasingly found to be helpful in building cases for BGI and these include measures to improve biodiversity, reduce pollution, increase carbon sequestration and combat urban heating.

Moreover, representing drainage systems in flood models is a challenging task, especially when data is unavailable. Future work is planned to improve the accuracy of the cost-benefit *'source-receptor'* framework by incorporating the sub-surface system into the model or developing new novel methodologies to accurately represent the sewer drainage network by generating synthetic inlets according to the study area and the design standards worldwide. These approaches will provide modellers with flexibility in cases where access to data is limited (Bertsch et al., 2017; Costabile et al., 2023; Dasallas et al., 2023; C. Iliadis et al., 2023; Singh et al., 2023).

5. Conclusions

This study demonstrates a novel approach to link surface water flooding information with the exposure and the flood damages to buildings in urban fabric and catchments. It uses a detailed hydrodynamic model to identify how many buildings are impacted by flooding during multiple magnitude storm events, the incurred damages and the potential locations to add the most effective type of BGI. The combination of hydrological data, flood dynamics and cost-benefits could guide spatial prioritization for intervention in critical locations. Furthermore, the proposed framework has all the necessary principles to become a standard planning tool for flood risk management due to the simplicity of every step in the proposed methodology and the quantitative exposure outputs from flood models at individual building level.

The systematic procedure of classifying the buildings at high and medium flood risk and calculating the damages by switching-off the rainfall in every cell for multiple storm events allows identification of areas and properties with high contribution and high direct flood impact. The proposed combination and comparison between the cost-benefit and the available green spaces provides information to choose different types of interventions according to the intensity of the storm, e.g. permeable pavements, water butts and green roofs for storms with low intensity and ponds and swales for storms with high intensity or even a combination of interventions according to flood results in the study area. Furthermore, this framework offers the flexibility to be applied in larger dense cities and catchments where the results offer more options for flood management intervention. For example, a target percentage of rainfall could be removed in every cell, for example 5% of the total storm, instead of unrealistic total capture when the rainfall is extremely high (e.g. more than 100 mm per hour).

Finally, much further work is planned to extend the capability of this cost-benefit *'source-receptor'* flood risk framework, such as automating the procedure to add BGI instead of manually investigating the best location in every cell (Rehman et al., 2023), and considering combinations of larger number of interventions with smaller footprints and lower unit cost to improve feasibility and flexibility of implementation.

**Acknowledgements:** This work has been funded by the Engineering and Physical Science Research Council (EPSRC) as part of the Centre for Doctoral Training in Water Infrastructure and Resilience (WIRe, EP/S023666/1).

**Conflicts of Interest:** The authors declare no conflict of interest.

**Supplementary material:** Appendix A - Exposure, Appendix B - Max Capture Flood Damages, Appendix C – Intervention